\documentstyle[twocolumn,graphics,prb,aps,floats]{revtex}
\begin{document}

\wideabs{
\draft
\title{Low-Energy Quasiparticles in Cuprate Superconductors: 
A Quantitative Analysis}

%
%
%
%

\author{May Chiao\cite{mayaddr},
R.W. Hill, Christian
Lupien 
and Louis Taillefer}

\address{Canadian Institute for Advanced Research}

\address{
Department of Physics, University of Toronto, Toronto,
Ontario, Canada M5S 1A7}

\author{P. Lambert and
R. Gagnon}

\address{Department of Physics, McGill University, Montr\'eal,
Qu\'ebec, Canada H3A 2T8}

\author{P. Fournier}

\address{Center for Superconductivity Research\\ Department of Physics, 
University of Maryland, College Park, MD 20742, USA}

\date{\today}

\maketitle

\begin{abstract}
  
A residual linear term
is observed in the thermal conductivity of
optimally-doped Bi$_2$Sr$_2$CaCu$_2$O$_8$ at very low temperatures
whose magnitude is 
in excellent agreement with the value expected from Fermi-liquid theory
and the $d$-wave energy spectrum measured by photoemission spectroscopy, 
with no adjustable parameters.
This solid basis allows us to make a quantitative analysis of thermodynamic properties
at low temperature and establish that thermally-excited quasiparticles are a significant,
perhaps even the dominant mechanism in suppressing the superfluid density in  
cuprate superconductors Bi$_2$Sr$_2$CaCu$_2$O$_8$ and YBa$_2$Cu$_3$O$_7$.

\end{abstract}

}

The superconducting order parameter of the
archetypal
high T$_c$ compounds YBa$_2$Cu$_3$O$_7$ and
Bi$_2$Sr$_2$CaCu$_2$O$_8$ is widely agreed to have $d$-wave symmetry,
yet there is no consensus on the correct theoretical description of 
their superconducting state properties, let alone those of the metallic state.
A fundamental issue in the current debate is the nature of the electronic
excitations in these systems, and whether long-lived quasiparticles exist \cite{Kaminski}
 or not.\cite{Valla}
Another debate concerns the dominant mechanism responsible for 
the thermal suppression of the superfluid density, 
whether it be 
$d$-wave nodal
quasiparticles \cite{WenLee} or phase fluctuations,\cite{Carlson} for example.
One way to shed new light on these issues is to go beyond the usual qualitative
temperature dependence of physical properties, and look closely at their
magnitude.  Our specific approach is to
examine quantitatively the basic thermodynamic and transport
properties of these two superconductors
within a Fermi-liquid description of $d$-wave quasiparticles, grounded
in a spectroscopic measurement of the energy spectrum,
and see whether a consistent 
description at low energies can be achieved.

The $d_{x^2-y^2}$ gap function goes to zero at four nodes along 
the $k_x=\pm k_y$ directions, producing a 
conelike quasiparticle excitation spectrum at low energies:

\begin{equation}
E~ =~ \hbar~\sqrt{~v_F^2~k_1^2~+~v_2^2~k_2^2}
\end{equation}

where $v_F$ and $v_2$ are the energy dispersions, or quasiparticle
velocities, along directions normal ($\parallel{\bf k}_1$) and
tangential ($\parallel{\bf k}_2$) to the Fermi surface,
at each node.
This spectrum is associated
with the two-dimensional CuO$_2$ plane that is the
fundamental
building block of all cuprates.  It neglects any possible
dispersion in the third direction (along the $c$-axis), as well as
excitations associated with the one-dimensional CuO chains found in
some crystal structures, notably in YBa$_2$Cu$_3$O$_{7-\delta}$
(along the $b$-axis).

This simple spectrum gives rise to a quasiparticle density of states
which is linear in energy:

\begin{equation}
N(E)~=~\frac{2}{\pi \hbar^2}~\frac{1}{v_F v_2}~E
\end{equation}

which in turn leads to a $T^2$ dependence of the electronic specific
heat and a linear $T$ dependence of the superfluid density, for example. 
In a realistic treatment, one needs to include the effect of impurity 
scattering and electron-electron interactions.
One usually accounts for the former in terms of a
single, isotropic scattering rate, parameterized by an
impurity bandwidth $\gamma$. At energies below $\gamma$, 
known as the ``dirty'' limit, one expects a profound modification
of the density of states which acquires a residual finite
value $N(0) \propto \gamma$.  At energies well above 
$\gamma$, in the
``clean'' limit, $N(E) \propto E$ and one recovers
many of the straightforward temperature dependences. 
Going beyond this, Durst and Lee recently included vertex 
corrections, which arise because of the anisotropy of scattering in a $d$-wave 
superconductor.\cite{Durst} 
The importance of Fermi-liquid corrections has also been emphasized, 
whereby electron-electron interactions renormalize the 
normal fluid density.\cite{WenLee,Durst,Millis}

In this paper, we use a measurement of the in-plane thermal
conductivity at very low temperature to extract a value for the ratio
$v_F/v_2$ in optimally-doped YBa$_2$Cu$_3$O$_{7-\delta}$ (YBCO) and
Bi$_2$Sr$_2$CaCu$_2$O$_8$ (BSCCO).  
We proceed to show first that for BSCCO this ratio is in excellent
agreement with the values of $v_F$ and $v_2$ measured separately
by
angle-resolved 
photoemission spectroscopy (ARPES).
We then use the ratio to calculate the drop in superfluid density within a Fermi-liquid
description and compare this with the experimental results obtained from penetration depth
measurements.
Finally, we extend our quantitative analysis to include specific heat measurements in YBCO.
From the overall analysis, we conclude that the superconducting state of the cuprates is
well described by Fermi-liquid theory, at least at low energy and optimal doping.\\

\paragraph*{Thermal conductivity}
The thermal conductivity of YBa$_2$Cu$_3$O$_{6.9}$ 
and Bi$_2$Sr$_2$CaCu$_2$O$_8$
was measured using a steady-state method,
described elsewhere.\cite{Taillefer}  The samples were single
crystals 
grown via standard flux techniques
and oxygenated so as to
obtain the maximum $T_c$ (optimal doping), quoted in Table I.
Both crystal structures are made of CuO$_2$ planes stacked along
the $c$-axis, with a density 30\% higher in YBCO, due to its lower
average interplane spacing, given in Table I.
The conductivity of YBCO was measured along the $a$-axis
in untwinned crystals, so
as to avoid the contribution of CuO chains.

\begin{table*}
\widetext
\caption{Comparison of YBa$_2$Cu$_3$O$_{6.9}$ and
Bi$_2$Sr$_2$CaCu$_2$O$_8$ at optimal doping. In YBCO, all directional
properties are for the $a$-axis (no chain contribution).
$d/n$ is the average separation between CuO$_2$ planes stacked along the
$c$-axis.
The zero-temperature
penetration depth $\lambda (0)$ was measured by far-infrared reflectivity
in YBCO \protect\cite{Basov} and by DC magnetisation in BSCCO. \protect\cite{Waldmann}
The Fermi velocity $v_F$ and momentum $k_F$ were obtained from angle-resolved photoemission.
\protect\cite{Schabel,Mesot}
$v_F/v_2$ is the ratio of quasiparticle velocities,
obtained via Eq.~(3), using
the residual linear term $\kappa_0/T$ measured in the thermal conductivity at $T \to 0$. 
$S = d \Delta(\phi) / d\phi = \hbar k_F v_2$ is the slope of the gap at the node calculated using 
$k_F$, $v_F$ and $v_F/v_2$. $\Delta_{\rm max}$ is the gap maximum as seen in $c$-axis tunneling
by STM,\protect\cite{Maggio,Wei,Renner} with $S = \mu \Delta_{\rm max}$.
The linear drop in superfluid density with temperature is expressed as 
$\lambda^2(0) d \lambda^{-2}/dT$, obtained from the penetration depth measured at microwave
frequencies.\protect\cite{UBC-LT21,Waldram}
$\alpha^2$ is the Fermi-liquid correction computed from Eq.~(7), using
the measured values of $d \lambda^{-2}/dT$ and $\kappa_0/T$.}

\vskip 0.5 cm
\begin{tabular}{lcccccccccccc}

\\
 & $\frac{d}{n}$ & $T_c$ & $\lambda (0)$ & $ v_F$ & $k_F$ & $\frac{\kappa_0}{T}$ 
& $\frac{v_F}{v_2}$ & $S$ & $\Delta_{\rm max}$ & $\mu$ & $\lambda^2(0) \frac{d \lambda^{-2}}{dT}$ & $\alpha^2$ \\
\\
      & (\AA) & (K)   &   (\AA)   & (km/s) & (\AA$^{-1})$ &  $(\frac{\rm mW}{\rm K^2 cm})$  
      &           & (meV) & (meV) &  & (K$^{-1}$) &                \\
\\
\hline
\\
YBCO  & 5.85  &  93.6 &    1600   &   $\sim 250$ &  $\sim 0.8$    &  0.14        
      &  14     &   94 & $\sim  20$   &  4.7  &   (205 K)$^{-1}$  &  0.46  \\
BSCCO  & 7.72  &  89 &    2100   &   $ 250$ &  $ 0.74$    &  0.15        
      &  19     &   64  &   $\sim 40$   &  1.6  &   (120 K)$^{-1}$ &   0.43  \\
\\
\end{tabular}
\narrowtext
\end{table*}

In Fig.~1 we present the low-temperature thermal conductivity, 
$\kappa$,
of BSCCO, and compare it with that of YBCO obtained previously.\cite{Chiao}
By plotting $\kappa/T$ vs $T^2$, we can separate the linear
quasiparticle term from the cubic phonon term (see 
Ref.\ \onlinecite{Taillefer}).  A finite residual linear term $\kappa_0/T$
(the value of $\kappa/T$ as $T\to$~0) is observed, of similar
magnitude for the two cuprates, given in Table I. (Note that
the value for YBCO is an average over several samples,\cite{Chiao}
only one of which is displayed in Fig.~1.)
The error bar on these numbers is approximately $\pm
20\%$, arising about equally from the uncertainty in the extrapolation and in
the geometric factor of each sample.

Calculations for the transport of heat by $d$-wave
quasiparticles in two dimensions give:\cite{Durst,Graf}

\begin{equation}
\frac{\kappa_{0}}{T}~ =~ \frac{k_B^2}{3\hbar} ~\frac{n}{d}
~(\frac{v_F}{v_2} + \frac{v_2}{v_F})~ \simeq~ 
\frac{k_B^2}{3\hbar}~ \frac{n}{d} ~
(\frac{v_F}{v_2})
\end{equation}

where $n/d$ is the stacking density of CuO$_2$ planes.
The residual conduction is due to a fluid of zero-energy
quasiparticles induced by the pair-breaking effect of impurity
scattering near the nodes in the gap, and it is independent of impurity
concentration.
This universal character of
$\kappa_0/T$ was demonstrated
explicitly for both YBCO \cite{Taillefer} and BSCCO. \cite{Kamran}

Durst and Lee recently showed Eq.~(3) to be valid even when vertex
and Fermi-liquid corrections are taken into account,\cite{Durst}
so that unlike charge transport, heat transport is {\it
  not} renormalized by either correction.  The universal
character and the absence of renormalization make thermal conductivity
a privileged probe of the quasiparticle spectrum in a $d$-wave
superconductor, providing a simple and direct measurement of $v_F/v_2$
in the cuprates.  From the measured $\kappa_0/T$ and the known values
of $n/d$ (see Table I), we obtain:

\begin{figure}
\resizebox{\linewidth}{!}{\includegraphics{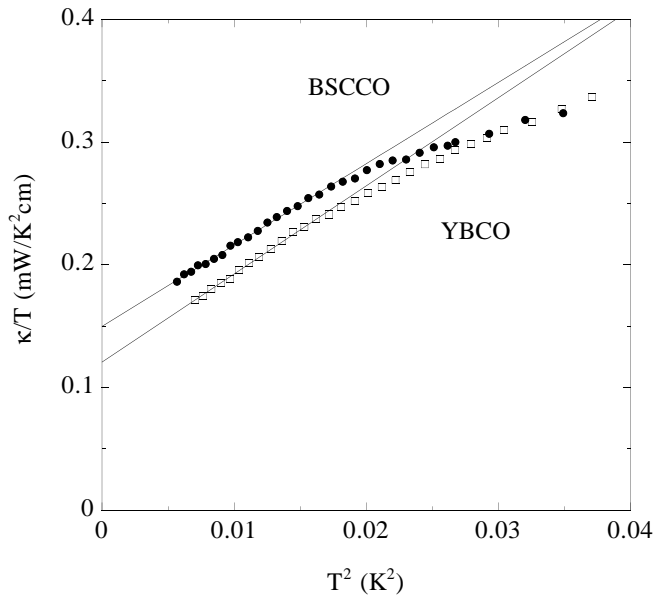}}
\caption{Thermal conductivity divided by temperature vs $T^2$ of
YBa$_2$Cu$_3$O$_7$ (squares) and Bi$_2$Sr$_2$CaCu$_2$O$_8$ (circles),
at optimum doping.  The lines are linear fits to the data below
130~mK, with extrapolated values given in Table I.}
\end{figure}

\begin{equation}
{\rm BSCCO~:}~~~~~\frac{v_F}{v_2} = 19 
\end{equation}

\begin{equation}
{\rm YBCO~~:}~~~~~\frac{v_F}{v_2} = 14
\end{equation}

with an uncertainty of about $\pm 20\%$. 
(Note that the ratio for YBCO is twice the value of $\sim 7$ often used in 
the literature.)\\

\paragraph*{ARPES}
Angle-resolved photoemission spectroscopy has established
the existence of a Fermi surface in YBCO \cite{Schabel} and 
BSCCO \cite{Kaminski,Valla} and revealed
directly the {\bf k}-dependence of the gap characteristic of 
$d_{x^2-y^2}$ symmetry.\cite{Shen,Mesot}
In BSCCO, the nodes are along the (0,0) to ($\pi$,$\pi$) direction, at 
$k = k_F = 0.74 ~{\rm \AA}^{-1}$,\cite{Mesot}
where the energy has a dispersion along ${\bf k}_1$
given by $v_F = 2.5 \times 10^7$ cm/s.\cite{Mesot}
As for the dispersion along ${\bf k}_2$ (or $\phi$), 
Mesot {\it et al}. \cite{Mesot} were recently able to
extract 
$S = |d\Delta/d\phi|_{node}$, the slope of the gap at the node.
For a crystal near optimal doping ($T_c$ = 87 K), they obtain
$S = 60 \pm 5$ meV ($= 1.7 \Delta_{\rm max}$, where $\Delta_{\rm max}$
is the gap maximum, at $\phi = 0$).
This yields $v_2 = S / \hbar k_F = 1.2
\times 10^6$ cm/s, so that $v_F/v_2 = 20$. 

A hotly debated question is whether the excitations in the vicinity of the Fermi surface,
in particular along the diagonals, can be treated as the usual Landau/BCS 
quasiparticles.\cite{Kaminski,Valla}
The excellent quantitative agreement we find between the spectroscopic
and the transport measurement of $v_F/v_2$ in BSCCO 
strongly validates 
a Fermi-liquid description of the superconducting state in cuprates, at least at low energies.

Unfortunately, 
ARPES measurements in YBCO 
have been less successful so far.
The Fermi surface is more complicated, with
bilayer splitting of the plane bands \cite{Schabel} and an added band
for the CuO chains. 
When averaged over the two plane bands,
the band crossing and dispersion
at the Fermi energy along the
(0,0) to $(\pi,\pi)$ direction
are close to those quoted above for BSCCO,
namely $v_F \simeq 2.5 \times 10^7$~cm/s and $k_F \simeq 0.8$~
\AA$^{-1}$,\cite{Schabel} albeit with greater uncertainty. 
The gap structure has not yet been resolved with sufficient resolution 
to provide a measurement of $v_2$. 
The thermal conductivity data may be used instead:
with $v_F \simeq 2.5 \times 10^7$~cm/s, Eq.~(5) yields
$v_2 \simeq 1.8 \times 10^6$~cm/s. This implies that
the slope of the gap at the node
in YBCO is 1.5 times {\it larger} than in BSCCO, with
$S = \hbar k_F v_2 \simeq 95$~meV,
in contrast with evidence from STM measurements of $c$-axis tunneling
that
the gap {\it maximum} in YBCO is
{\it smaller} than in BSCCO, namely $\Delta_{\rm max} \simeq 20$~ meV \cite{Maggio,Wei} vs 
$\simeq 40$~ meV 
\cite{Renner}. This suggests a strikingly different angular dependence of the gap
function, with a ratio of slope to gap maximum 3 times larger in YBCO (see Table I),
under the assumption that $v_F$ is the same in both materials.\\

\paragraph*{Superfluid density}
As the temperature is increased from $T=0$,
the thermal excitation of nodal quasiparticles
causes the normal fluid 
density, $\rho_n(T)$, to grow linearly with temperature.
In the clean limit, at low temperature:\cite{WenLee,Durst,Millis,Xu}

\begin{equation}
\frac{\rho_n(T)}{m}
~=~\frac{2{\rm ln}2}{\pi}~\frac{k_B}{\hbar^2}~
        \frac{n}{d} ~ \alpha^2~ (\frac{v_F}{v_2})~T
\end{equation}

where $m$ is the mass of the carriers
and 
$\alpha^2$
is the Fermi-liquid correction for charge currents.\cite{Alpha}

A linear temperature dependence of $\rho_n(T)$
is a characteristic feature of
most cuprate superconductors, as
revealed through measurements of the penetration depth $\lambda (T)$, 
via the relation $\rho_s(T)/m = \rho_s(0)/m - \rho_n(T)/m = 
c^2/4\pi e^2 \lambda^2(T)$.
From the data of Hardy and co-workers on 
untwinned crystals of YBCO \cite{Zhang,UBC-LT21} --
again taking the $a$-axis results to avoid chain contributions -- and
from measurements by Waldram and co-workers in BSCCO,\cite{Waldram}
one finds the 
slope of $\lambda^2(0)/\lambda^2(T)$ at low temperatures, given
in Table I for
optimal doping.  
Combining Eqs.~(3) and (6), we can then solve for the
Fermi-liquid correction, via:

\begin{equation}
\frac{d \lambda^{-2}(T)}{d T} =  - 2.93 \times 10^{13}
~ \frac{\kappa_0}{T} ~ \alpha^2
\end{equation}

with $\lambda$ in meters and $\kappa_0/T$ in W~K$^{-2}$~m$^{-1}$.
Using the values for $\lambda (0)$ quoted in Table I,
we get:

\begin{equation}
{\rm BSCCO~:}~~~~~\alpha^2~=~0.43
\end{equation}

\begin{equation}
{\rm ~YBCO~:}~~~~~\alpha^2~=~0.46
\end{equation}

\begin{figure}
\resizebox{\linewidth}{!}{\includegraphics{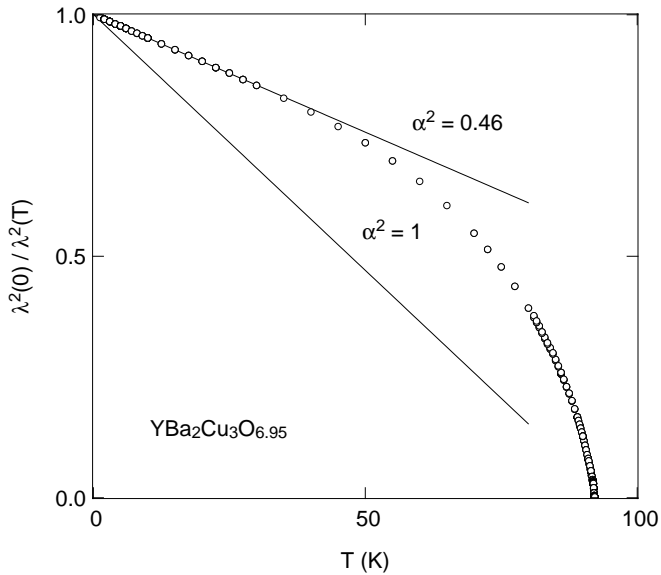}}
\caption{Temperature dependence of the superfluid density, normalized to unity at
$T=0$, for optimally-doped YBCO ($a$-axis) (from Ref.\ \protect\onlinecite{UBC-LT21}). The lines are the expected low temperature behaviour
calculated from Eq.~(7) with ($\alpha^2 \neq 1$) and without ($\alpha^2=1$) Fermi-liquid
interactions.}
\end{figure}

In other words, the observed drop in superfluid density is about 2 times
weaker than expected from a calculation neglecting interactions, as shown graphically
for YBCO in Fig.~2.
Since electrons in cuprates are highly correlated, 
a renormalization by
a factor of 2 seems entirely
plausible. The fact that it is comparable in the two
compounds is not unexpected, given that the  Fermi velocities,
themselves renormalized by interactions,\cite{Millis} are comparable.
We stress that an estimate of the Fermi-liquid correction to $\rho_s(T)$
does not require a separate knowledge of $v_F$ and $v_2$, and heat conduction
-- unlike heat capacity, for example --
can provide directly the appropriate combination of the two parameters, {\it i.e.}
their ratio.
Note that in the case of BSCCO, Mesot {\it et al}. \cite{Mesot} were the first
to report
an estimate of the renormalization factor,
based on their ARPES data. (The fact that they obtain a slightly different value,
namely $\alpha^2=0.32$, is due to their use of different penetration depth data, which we consider
to be less reliable because restricted to temperatures above 17~K.)

Given the numbers that emerge from the analysis, 
it seems fair to conclude that
the thermal excitation of quasiparticles is a significant, perhaps even the dominant
mechanism in suppressing
the superfluid density of these two 
cuprate superconductors. 
It is interesting that electron-electron interactions 
appear to be such as to weaken
this process.

It is perhaps worth noticing that although the density of superfluid at $T=0$
($\propto \lambda^{-2}(0)$) is 1.7 times higher in YBCO, the normal fluid density grows at exactly the same rate
in both compounds at low temperature. This would naively suggest that T$_c$ should be much higher
in YBCO, while it is in fact not very different (5\% higher). This is because $\rho_s(T)$
acquires a much stronger downward curvature near T$_c$ in YBCO.
Therefore, a major difference must develop at higher temperatures. Part of the answer must come from
the very different curvature of the gap function away from the nodes, at high energies, as parametrized
by $\mu = S / \Delta_{\rm max}$ being much larger in YBCO (see Table I).\\

\paragraph*{Specific heat}
We complete our quantitative analysis by looking at the electronic specific heat,
which is simply derived from Eq.~(2), in the clean limit:

\begin{equation}
C_{el}(T) = \frac{18 \zeta(3)}{\pi}~
\frac{k_B^3}{\hbar^2}~ \frac{n}{d}~ (\frac{1}{v_F v_2})~T^2
\end{equation}

where $\zeta(3) \simeq 1.20$.
Extracting this electronic contribution
from the total specific heat has been a controversial exercise, since the data can be
fitted equally well without a $T^2$ term.  
In YBCO at optimal doping, the value quoted in the literature is
0.1~mJ~K$^{-3}$~mole$^{-1}$,\cite{Moler,Wright} albeit with a $\pm$~60\%
uncertainty.
Using $v_F = 2.5 \times 10^7$~cm/s and
$v_2
=1.8 \times 10^6$~cm/s,
Eq.~(10) gives 0.065~mJ~K$^{-3}$~mole$^{-1}$ -- a value within the 
experimental uncertainty.

An alternative approach is to extract 
$v_2$ from the field dependence of the specific heat.
The Doppler shift of
quasiparticle states near the nodes in the presence of the superfluid flow
around vortices leads
to an increase in the specific heat proportional to $\sqrt{H}$.\cite{Volovik}  In terms
of the nodal spectrum, the magnitude of the effect is related only to the
slope of the gap at the node, $v_2$.  In the clean limit, the electronic specific
heat of CuO$_2$ planes (per unit volume), calculated by averaging the effect of
the Doppler shift over a single vortex-lattice unit cell, is given by:\cite{Kubert}

\begin{equation} \frac{C_{el}}{T} = \frac{4 k_B^2}{3
\hbar}~ \sqrt{\frac{\pi}{\Phi_0}}~ \frac{n}{d}~ (\frac{a}{v_2})~~ \sqrt{H}
\end{equation}

where $\Phi_0$ is the flux quantum, and $a$ is a vortex-lattice parameter
of order unity. 
Such a $\sqrt{H}$ dependence has been seen in measurements
on YBCO \cite{Moler,Wright} and the coefficient, of magnitude 
0.9~mJ~K$^{-2}$~mole$^{-1}$~T$^{-1/2}$, determined with
greater accuracy ($\pm10\%$) than the corresponding zero-field $T^2$
term.  
Using $v_2 =1.8 \times 10^6$~cm/s and $a=1$ in Eq.~(11) gives
0.6~mJ~K$^{-2}$~mole$^{-1}$~T$^{-1/2}$.

It is clear that both aspects of the specific heat data for YBCO are in
reasonable quantitative agreement with our thermal conductivity result for
$v_F/v_2$
and the value of $v_F$ from ARPES. In a refined  
treatment, one would take into account the contribution of CuO chains
to the density of states  
(and hence to the specific heat).
In this respect, it is interesting that Junod and
co-workers extract $T^2$ and $\sqrt{H}$ coefficients 
for an {\it overdoped} crystal of YBCO which
are somewhat larger:\cite{Junod} 0.20$\pm 0.05$~mJ~K$^{-3}$~mole$^{-1}$ and
1.3$\pm 0.1$ ~mJ~K$^{-2}$~mole$^{-1}$~T$^{-1/2}$.  It is not unreasonable to
attribute this increase to a larger chain density of states, such as
would explain the decreasing $a-b$ anisotropy in the linear
temperature drop of the superfluid density observed in crystals of
YBCO as one moves from overdoped to underdoped.\cite{UBC-LT21}\\

In conclusion, we have provided a quantitative analysis of low-temperature
data
for the cuprate superconductors YBCO and BSCCO at optimal doping
which compared
results from our thermal
conductivity measurements with existing results from
ARPES, microwave penetration depth and specific heat.
Within a Fermi-liquid
model 
of $d$-wave quasiparticle excitations with
interactions, we find all data consistent
with a single set of parameters. 
The Fermi velocity 
and the Fermi-liquid renormalization of charge currents are found to be
roughly the same
in both compounds (as is T$_c$), with
$v_F \simeq 2.5 \times 10^7~{\rm cm/s}$ and
$\alpha^2~\simeq~0.4-0.5$, whereas the slope of the gap at the node
is about 1.5 times steeper in YBCO.

In particular, the thermal excitation of quasiparticles emerges as a sufficient
mechanism for suppressing the superfluid, and there is no clear evidence for
a significant contribution from phase fluctuations at low temperature, 
at least at optimal
doping. The success of a Fermi-liquid description for the low-temperature properties
should not be taken to mean that the normal state of cuprates is a Fermi liquid. Nor
should it be viewed as supporting a BCS theory of the superconducting state, given
the fact that, for example, the thermal excitation of quasiparticles does not have the expected
impact on the gap itself,\cite{WenLee}
which remains undiminished up to high temperatures.\cite{Renner}\\


We are grateful to P.A. Lee, M.R. Norman, A. Junod and I. Vekhter 
for stimulating
discussions.  LT, MC and CL are most grateful for the generous
hospitality they received at the Grenoble High Magnetic Field
Laboratory and the Centre de Recherches sur les Tr\`es Basses
Temp\'eratures in Grenoble, France, where this article was written.
This work was supported by the Canadian Institute for Advanced
Research and funded by NSERC of Canada.  MC acknowledges support from
the Carl Reinhardt Foundation.  CL and LT respectively ackowledge 
the support of a Post-Graduate Scholarship and a
Steacie Fellowship from NSERC.
The work in Maryland was supported by the NSF Division of Condensed 
Matter Physics under grant No. DMR 9732736.\\
\\
\\


\begin{references}

\bibitem[*]{mayaddr} 
Present address: Cavendish Laboratory, University of
Cambridge, Madingley Road, Cambridge CB3 0HE, UK.

\bibitem{Kaminski} A. Kaminski, J. Mesot, H.M. Fretwell, J.C. Campuzano,
M.R. Norman, M. Randeria, H. Ding, T. Sato, T. Takahashi, T. Mochiku, K. Kadowaki,
and H. Hoechst, Phys. Rev. Lett. {\bf 84}, 1788 (2000).

\bibitem{Valla} T. Valla, A.V. Federov, P.D. Johnson, B.O. Wells, S.L. Hulbert,
Q. Li, G.D. Gu, and N. Koshizuka, Science {\bf 285}, 2110 (1999).

\bibitem{WenLee} X.-G. Wen and P.A. Lee, Phys. Rev. Lett. {\bf
80}, 2193 (1998).

\bibitem{Carlson} E.W. Carlson, S.A. Kivelson, V.J. Emery,
and E. Manousakis, Phys. Rev. Lett. {\bf 83},
612 (1999).

\bibitem{Durst} A.C. Durst and P.A. Lee, to appear in Phys. Rev. B (preprint,
http://xxx.lanl.gov/abs/cond-mat/9908182).

\bibitem{Millis} A.J. Millis, S.M. Girvin, L.B. Ioffe, and
A.I. Larkin, J. Phys. Chem. Solids {\bf 59},
1742 (1998).

\bibitem{Taillefer} Louis Taillefer, Benoit Lussier, Robert Gagnon,
Kamran Behnia, and Herv\'e Aubin, Phys. Rev. Lett. {\bf
79}, 483 (1997).

\bibitem{Chiao} May Chiao. R.W. Hill, Christian Lupien, Bojana Popic,
Robert Gagnon, and Louis Taillefer, Phys. Rev. Lett. {\bf 82}, 2943
(1999).

\bibitem{Graf} M.J. Graf, S-K. Yip, J.A. Sauls, and D. Rainer, 
Phys. Rev. B {\bf 53}, 15147 (1996).

\bibitem{Kamran} K. Behnia, S. Belin, H. Aubin, F. Rullier-Albenque, S. Ooi,
T. Tamegai, A. Deluzet, and P. Batail, Proceedings of MOS'99 (Stockholm, 1999).

\bibitem{Schabel} M.C. Schabel, C.-H. Park, A. Matsuura, Z.-X. Shen,
D.A. Bonn, Ruixing Liang, and W.N. Hardy, Phys. Rev. B {\bf 57}, 6090
(1998).  

\bibitem{Shen} Z.-X. Shen, D.S Dessau, B.O. Wells, D.M. King, W.E. Spicer,
A.J. Arko, D. Marshall, L.W. Lombardo, A. Kapitulnik, P. Dickinson, and S. Doniach, 
Phys. Rev. Lett. {\bf
70}, 1553 (1993).

\bibitem{Mesot} J. Mesot, M.R. Norman, H. Ding, M. Randeria, J.C. Campuzano, 
A. Paramekanti, H.M. Fretwell, A. Kaminski, T. Takeuchi, T. Yokoya, T. Sato, 
T. Takahashi, T. Mochiku, and K. Kadowaki, Phys. Rev. Lett. {\bf 83}, 840
(1999). 

\bibitem{Maggio} I. Maggio-Aprile, Ch. Renner, A. Erb, E. Walker, and
O. Fischer, Phys. Rev. Lett. {\bf 75}, 2754
(1995).

\bibitem{Wei} J.Y.T. Wei, N.-C. Yeh, D.F. Garrigus, and M. Strasik, 
Phys. Rev. Lett. {\bf
81}, 2542 (1998).

\bibitem{Renner} Ch. Renner, B. Revaz, J.-Y. Genoud, K. Kadowaki,
and O. Fischer, Phys. Rev. Lett. {\bf 80},
149 (1998).

\bibitem{Xu} D. Xu, S. K. Yip, and J.A. Sauls, Phys. Rev. B {\bf 51},
16233 (1995).

\bibitem{Alpha} In the isotropic Fermi-liquid theory, $\alpha = 
1+\frac{F_1^s}{2}$, where $F_1^s$ is the $l=1$ spin-symmetric Landau 
parameter.\cite{Durst}

\bibitem{Zhang} Kuan Zhang, D.A. Bonn, S. Kamal, Ruixing Liang, D.J. Baar,
W.N. Hardy, D. Basov, and T. Timusk, Phys. Rev. Lett. {\bf 73}, 2484
(1994).

\bibitem{UBC-LT21} D.A. Bonn, S. Kamal, A. Bonakdarpour, Ruixing Liang, W.N. Hardy,
C.C. Homes, D.N. Basov, and T. Timusk, Czech. J. Phys. {\bf 46}, 3195
(1996).

\bibitem{Waldram} Shih-Fu Lee, D.C. Morgan, R.J. Ormeno, D.M. Broun, R.A. Doyle,
J.R. Waldram, and K. Kadowaki, Phys. Rev. Lett. {\bf 77}, 735
(1996).  

\bibitem{Basov} D.N. Basov, R. Liang, D.A. Bonn, W.N. Hardy, B. Dabrowski, M. Quijada, 
D.B. Tanner, J.P. Rice, D.M. Ginsberg, and T. Timusk, Phys. Rev. Lett. {\bf 74}, 598
(1995).

\bibitem{Waldmann} Waldmann, F. Steinmeyer, P. M\"uller, J.J. Neumeier,
F.X. R\'egi, H. Savary, and J. Schnek, Phys. Rev. B. {\bf 53}, 11825
(1996).

\bibitem{Moler} K.A. Moler, D.L. Sisson, J.S. Urbach, M.R. Beasley,
A. Kapitulnik, D.J. Baar, R. Liang, and W.N. Hardy, Phys. Rev. B {\bf 55}, 3954
(1997).

\bibitem{Wright} D.A. Wright, J.P. Emerson, B.F. Woodfield, J.E. Gordon,
R.A. Fisher, and N.E. Phillips, Phys. Rev. Lett. {\bf 82}, 1550
(1999).

\bibitem{Volovik} G. Volovik, J.E.T.P. Lett. {\bf 58}, 469 (1993).

\bibitem{Kubert} C. K\"ubert and P.J. Hirschfeld,
Sol. St. Commun. {\bf 105}, 459 (1998).

\bibitem{Junod} A. Junod, B. Revaz, Y. Wang, and A. Erb, 
Proceedings of LT22 (Helsinki, August 1999).

\end {references}
\end{document}